\documentclass[twocolumn,showpacs,preprintnumbers,amsmath,amssymb]{revtex4}

\usepackage{graphicx}
\usepackage{dcolumn}
\usepackage{bm}

\begin{document}


\title{Temperature Dependence of Zero-Bias Resistances \\
of a Single Resistance-Shunted Josephson Junction}

\author{Naoki  Kimura}
 \email{kimura@a-phys.eng.osaka-cu.ac.jp}
\author{Takeo  Kato}%
\affiliation{
Department of Applied Physics, Osaka City University,
3-3-138 Sugimoto Sumiyoshi-ku, Osaka 558-8585, Japan}
\date{\today}

\begin{abstract}
Zero-bias resistances of a single resistance-shunted Josephson junction
are calculated as a function of the temperature by means of 
the path-integral Monte Carlo method in case a charging energy $E_{\rm C}$
is comparable with a Josephson energy $E_{\rm J}$.
The low-temperature behavior of the zero-bias
resistance changes around $\alpha=R_{\rm Q}/R_{\rm S}=1$, where 
$R_{\rm S}$ is a shunt resistance and $R_{\rm Q}=h/(2e)^2$. 
The temperature dependence of the zero-bias 
resistance shows a power-law-like behavior whose exponent depends on
$E_{\rm J}/E_{\rm C}$. These results are compared with the experiments on
resistance-shunted Josephson junctions.
\end{abstract}

\maketitle

Dissipative quantum systems have been studied for several decades 
in various systems~\cite{Weiss99}. Among them, resistance-shunted 
Josephson junctions have been studied intensively as a typical 
system~\cite{Caldeira81,Schon90}.
Theoretically, it has been shown that a single resistance-shunted 
Josephson junction exhibits a dissipation-induced
superconductor-insulator transition at $\alpha=R_{\rm Q}/R_{\rm S}=1$, 
where $R_{\rm Q}=h/4e^2$ and $R_{\rm S}$ is a shunt
resistance~\cite{Schmid83,Fisher85}. This transition is studied
for $E_{\rm J}/E_{\rm C} \ll 1$ and $E_{\rm J}/E_{\rm C} \gg 1$,
and is also expected to occur at any ratio $E_{\rm J}/E_{\rm C}$, 
where $E_{\rm J}$ and $E_{\rm C}$ are a Josephson energy and 
charging energy of a junction, respectively. Recently, the phase diagram
has been obtained experimentally for a single 
resistance-shunted Josephson junction in the 
$\alpha-(E_{\rm J}/E_{\rm C})$ plane~\cite{Yagi97,Pentila99}.
Subsequently, the phase diagram has been obtained also for 
one-dimensional~\cite{Miyazaki02} and two-dimensional~\cite{Takahide00} 
arrays of resistance-shunted Josephson junctions. For these 
higher-dimensional systems, the corresponding phase diagrams are 
modified from the one for a single junction.

In this report, we study the temperature dependence of the zero-bias
resistance $R_0$. Analytical calculation has shown 
that $R_0$ behaves as $T^{2\alpha -2}$ for 
$\alpha > 1$ and $E_{\rm J}/E_{\rm C} \gg 1$~\cite{Weiss85}. 
The renormalization group approach have indicated that 
also for any ratio of $E_{\rm J}/E_{\rm C}$, below a crossover temperature
$T_{\rm cr}$, the behavior $R_0\propto T^{2\alpha -2}$ may be 
observed~\cite{Fisher85}. However, there may be deviation from this
result for $E_{\rm J} \sim E_{\rm C}$ above $T_{\rm cr}$. 
In order to clarify this deviation, we study the zero-bias resistance 
by means of the path-integral Monte Carlo method. Similar numerical 
study of a single resistance-shunted Josephson junction has been
performed by Herrero {\it et.al.}~\cite{Herrero02}. They have 
discussed phase fluctuations along
the imaginary time, clarifying the superconductor-insulator 
phase boundary lies on $\alpha =1$ even for $E_{\rm J}\sim E_{\rm C}$. 
They have also obtained
the linear resistance on the Matsubara frequency, $R({\rm i}\omega_n)$.
They, however, have not obtained the zero-bias resistance on the real 
frequencies, which can be directly related to the experimental data.
In this paper, we calculate $R({\rm i}\omega_n)$ first, and obtain
the zero-bias resistance by extrapolation as
\begin{equation}
R_0 = \lim_{\omega_n \rightarrow 0} R({\rm i}\omega_n).
\label{eq:extra}
\end{equation}
We also compare the numerical results with the experiments
of resistance-shunted Josephson junctions.


The equation of motion of a single resistance-shunted Josephson junction 
is expressed by a phase difference $\phi$ as
\begin{equation}
\left(\frac{\Phi_0}{2\pi}\right)^2C \ddot{\phi}+
\left(\frac{\Phi_0}{2\pi}\right)^2\frac{1}{R_{\rm S}}\dot{\phi}+
\frac{\partial V}{\partial \phi} = 0,
\end{equation}
where $C$ is a capacitance, $R_{\rm S}$ is a shunt resistance, 
and $\Phi_0=h/2e$ is the flux quantum. 
The potential energy is given as
\begin{equation}
V(\phi) = - E_{\rm J}\cos \phi - I \frac{\Phi_0}{2\pi} \phi,
\end{equation}
where $I$ is an external current through the junction.  
This equation of motion is equivalent to a dissipative 
classical motion described by
\begin{equation}
M\ddot{\phi} + M\gamma \dot{\phi} + \frac{\partial V}{\partial \phi} = 0.
\end{equation}
Here, a particle mass $M$ corresponds to $\hbar^2/8E_{\rm C}$, and
a damping frequency $\gamma$ to $1/R_{\rm S}C$, where $E_{\rm C}=e^2/2C$.
In order to consider the dissipation effect, we use the Caldeira-Leggett 
model~\cite{Caldeira81}, in which, a reservoir consisting of 
infinite harmonic oscillators is coupled
to the system. In the path-integral formalism,
the partition function $Z$ is calculated by 
integrating the degrees of freedom of a reservoir as~\cite{Weiss99}
\begin{eqnarray}
Z &=& \int {\cal D} \phi (\tau) e^{-S_{\rm eff}/\hbar}, 
\label{eq:partition} \\
\nonumber S_{\rm eff} &=& \int_{0}^{\beta \hbar} d\tau \left[
\frac{\hbar^2}{16E_{\rm C}}\dot{\phi}(\tau)^2 + V(\phi (\tau)) \right]\\
&+& \frac{1}{2} \int_{0}^{\beta \hbar }d\tau \int_{0}^{\beta \hbar } d\tau'
\phi (\tau)K(\tau-\tau')\phi (\tau')
\end{eqnarray}
where $K(\tau)$ is a damping kernel given by
\begin{equation}
K(\tau) = \sum_{n} \frac{|\nu_n|}{R_{\rm S}C} e^{-i\nu_n\tau}.
\end{equation}
Here $\nu_n=2\pi n/\beta \hbar$ is the Matsubara frequency.

In order to apply the path-integral Monte Carlo method,
we introduce a discrete imaginary time $\tau_j = \hbar \beta j/P$,
where $P$ is a Trotter number assumed to be even 
and $j=0, 1, 2, \cdots, P-1$.
Then, the path $\phi(\tau)$ is approximated by a discrete path
$\phi_j = \phi(\tau_j)$. After the Fourier transformation 
$\phi_j = \overline{\phi} + 2\sum^{P/2-1}_{k=1}
(a_k\cos\frac{2\pi kj}{P} + b_k\sin\frac{2\pi kj}{P}) + a_{P/2} (-1)^j$, 
the partition function (\ref{eq:partition}) is approximately given as
\begin{eqnarray}
Z_P &=& A\int d\overline{\phi}\int \prod_{k=1}^{P/2-1}da_k db_k
\int da_{P/2} \ e^{- S_{{\rm eff},P}}, \\
\nonumber S_{{\rm eff},P} &=& - \sum_{k=1}^{P/2-1} 
\left( \frac{\beta E_CP^2}{2}
\sin^2\frac{\pi k}{P}+\alpha k \right) (a_k^2+b_k^2) \\
&+& \left( \frac{\beta E_CP^2}{2} + \frac{\alpha P}{2}\right) a_{P/2}^2 
+ \frac{\beta}{P}\sum_{l} V(\phi_l), 
\end{eqnarray}
where $\alpha = R_Q/R_S$ and $A$ is a constant. 
By this modification, the dissipative quantum mechanics is reduced to
a classical statistical problem. 
Applying the ordinary Monte Carlo method to this effective action, 
we can simulate the dissipative quantum system. Using the linear 
response theory, the zero-bias resistance can be related to the
correlation function
\begin{equation}
\langle \phi \phi \rangle_{\omega_n} = \int_0^{\hbar \beta}
d\tau \ e^{{\rm i}\omega_n \tau} \langle \phi(0) \phi(\tau) \rangle
\end{equation}
as~\cite{Schmid83}
\begin{equation}
R({\rm i}\omega_n)/R_{\rm Q} = \eta | \omega_n | \langle \phi \phi 
\rangle_{\omega_n},
\end{equation}
where $\eta = \alpha/2\pi$.  The zero-bias resistance is 
then obtained by the extrapolation (\ref{eq:extra}).

\begin{figure}[htbp]
\begin{center}
\includegraphics[width=7cm]{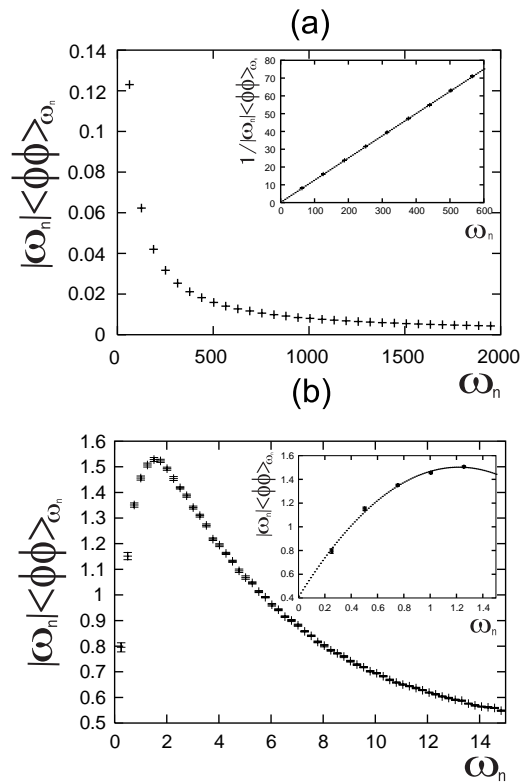}
\end{center}
\caption{Frequency dependence of the Matsubara resistance (a) 
for the high temperature region ($\beta=0.1$), and (b) for the low 
temperature region ($\beta=25.0$). We take $\alpha$ as $1.5$ and 
the ratio $E_{\rm J}/E_{\rm C}$ as $1$. The insets show 
the fitting procedure (see also the text).}
\label{fig:fitting}
\end{figure}

Generally, an analytic continuation of the correlation function
to the real frequencies may have a large noise, and is ill-defined.
Additionally in the present system, the correlation function is
not analytic at $\omega = 0$ due to the presence of
a characteristic factor $|\omega_n|$ of the ohmic damping.
Therefore, the Pad\'e approximation, which is one of the standard methods
of the analytic continuation, does not work well except for moderate
temperatures. Here, since we are interested only in $R({\rm i}\omega_n = 0)$,
we extrapolate the numerical data toward $\omega_n = 0$ by assuming 
a fitting function. At high temperatures, the numerical data of 
$R({\rm i}\omega_n)$ can be fitted well to the form
\begin{equation}
 R({\rm i}\omega_n)= \frac{1}{A|\omega_n|+B}. 
 \label{eq:fitting1}
\end{equation}
One example is shown in Fig.~\ref{fig:fitting}~(a). At low temperatures
($\beta > 5$), the deviation from the form (\ref{eq:fitting1}) is
observed. Then, we fit the data to the form
\begin{equation}
 R({\rm i}\omega_n) = A' + B'|\omega_n| + C' \omega_n^2 
\end{equation}  
by using the lowest five frequencies. One example of this case is shown in 
Fig.~\ref{fig:fitting}~(b). For moderate temperatures, 
$R_0$ is consistent with the one obtained by the Pad\'e approximation within
10 percent. We expect that the method adopted here for the extrapolation
is accurate enough to discuss the temperature dependence of $R_0$.

\begin{figure}[htbp]
\begin{center}
\includegraphics[width=7cm]{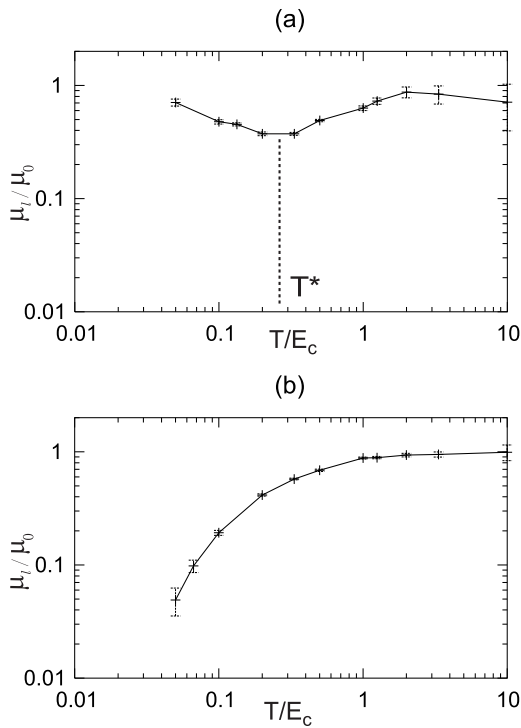} \\
\end{center}
\caption{Temperature dependence of the scaled linear mobility 
for (a) $\alpha=0.5$ and (b) $\alpha=2.0$. The ratio $E_J/E_C$
is taken as $1$. In case of $\alpha=0.5$, the linear mobility has 
a minimum value at $T = T^*$.}
\label{fig:paper1}
\end{figure}

Before showing the result of the zero-bias resistance $R_0$,
we study a linear mobility defined as $\mu_{\rm l}= \mu_0 R_0/R_{\rm Q}$, 
where $\mu_0=1/\eta$ is a linear mobility of a free particle.
The temperature dependence of the linear mobility 
has been studied by Fisher and Zwerger~\cite{Fisher85}
for $E_{\rm J}/E_{\rm C} \ll 1$.
They have showed that for $\alpha <1$, 
as the temperature decreases, the scaled linear mobility $\mu_{\rm l}/\mu_0$
shows a reentrant behavior; above a temperature $T^*$ it once decreases 
from $1$, and increases toward $1$ below $T^*$.
They have also claimed that for $\alpha >1$ the linear mobility decreases
monotonically as the temperature decreases, and becomes zero
at the zero temperature. The temperature dependence
of $\mu_{\rm l}/\mu_0$ in the regime $E_{\rm J}\sim E_{\rm C}$ has, 
however, not yet studied. The temperature dependence of $\mu_{\rm l}/\mu_0$
obtained in our simulation is shown in Fig.~\ref{fig:paper1} 
for $E_{\rm J}/E_{\rm C} =1.0$. These results show that the linear mobility 
behaves as predicted by Fisher and Zwerger even for $E_{\rm J}\sim E_{\rm C}$.
The similar behaviors are observed also for other values of 
$E_{\rm J}/E_{\rm C}$.

\begin{figure}[htbp]
\begin{center}
\includegraphics[width=7cm]{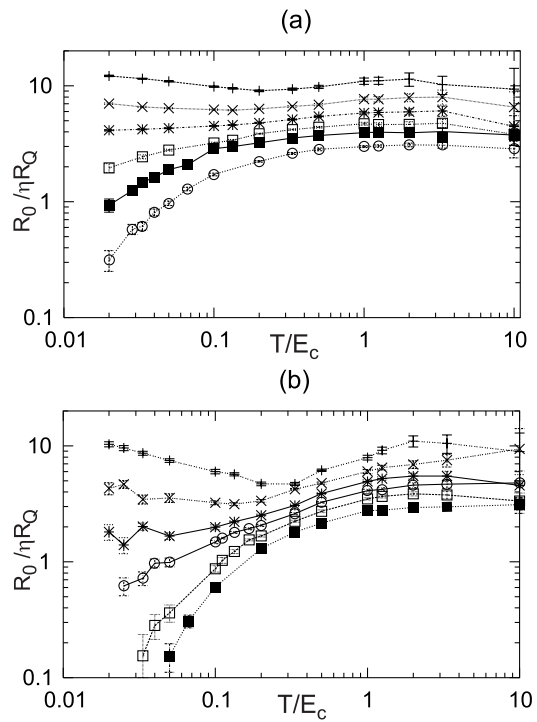}
\end{center}
\caption{Zero-bias resistance for various $\alpha$ for (a) 
$E_{\rm J}/E_{\rm C}=0.5$ and (b) $E_{\rm J}/E_{\rm C}=1.0$. 
From the top to the bottom, 
$\alpha$ is taken as $0.5, 0.75, 1.0, 1.25, 1.5$ and 
$2.0$.}
\label{fig:paper2}
\end{figure}

Next, the superconductor-insulator phase transition is studied
through the temperature dependence of the zero-bias resistance $R_0$.
We show $R_0$ as a function of the temperature for various values
of $\alpha$ in Fig.~\ref{fig:paper2}. Due to the difficulty of the
extrapolation, the temperature region in which $R_0$ is properly
obtained, is limited. Particularly, the lack of data at low temperatures
makes it difficult to distinguish the phase of the ground state. 
The qualitative feature, however, can be observed in these figures; 
For $\alpha < 1$, the result shows the tendency of a reentrant 
behavior, while for $\alpha > 1$ the result shows a monotonic decrease
as the temperature decreases. Hence, we expect that the phase
boundary lies on $\alpha =R_{\rm Q}/R_{\rm S}= 1$ even for $E_{\rm J}
\sim E_{\rm C}$.

\begin{figure}[htbp]
\begin{center}
\includegraphics[width=7cm]{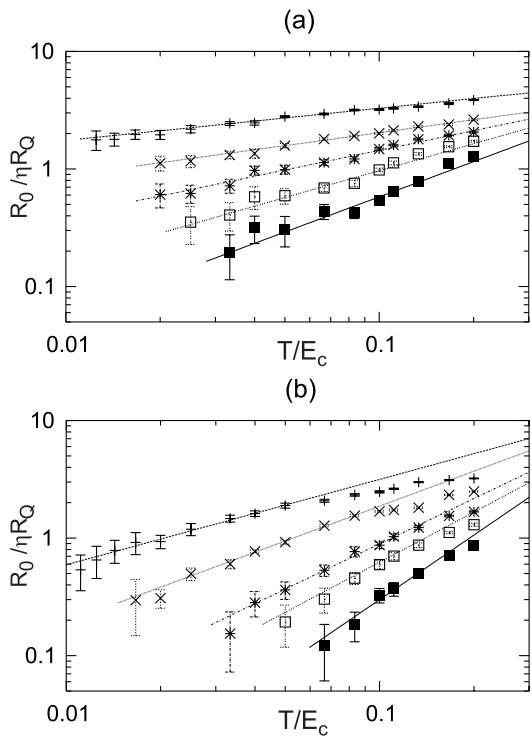}
\end{center}
\caption{Power-law-like behavior of the zero-bias resistances
for various $E_J/E_C$ values at (a) $\alpha=1.25$ and (b) $1.5$.
From the top to the bottom: 
$E_{\rm J}/E_{\rm C}=0.5,0.8,1.0,1.2$ and $1.5$.}
\label{fig:paper3}
\end{figure}

\begin{figure}[htbp]
\begin{center}
\includegraphics[width=7cm]{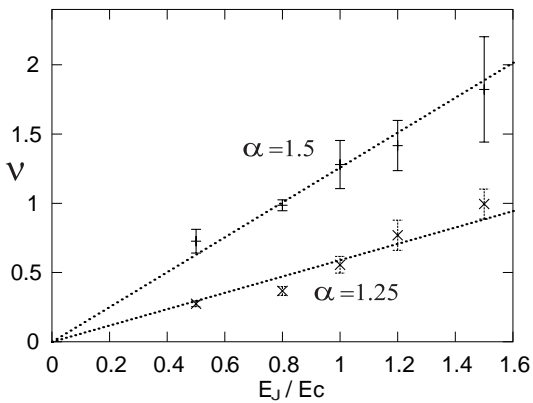}
\end{center}
\caption{Plot of the ratio $E_{\rm J}/E_{\rm C}$ vs. the exponent $\nu$.}
\label{fig:paper4}
\end{figure}

Finally, we discuss the temperature dependence of $R_0$ in
the superconducting phase ($\alpha > 1$). For both $\alpha = 1.25$
and $\alpha = 1.5$, the zero-bias resistances show a power-law-like
behavior ($R_0 \propto T^{\nu}$) in the low temperature region as shown 
in Fig.~\ref{fig:paper3}. This result disagrees 
with the analytic result ($R_0 \propto T^{2\alpha - 2}$) expected
in the low temperature limit;
The exponent $\nu$ seems to depend also on $E_{\rm J}/E_{\rm C}$. 
This indicates that above a crossover temperature $T_{\rm cr}$
there exists deviations from the result expected in
the renormalization group analysis, and that $T_{\rm cr}$ is
much smaller than the temperature range studied in this report.
Note that the exponent $\nu$ may change depending on the temperature 
range, and that $\nu$ should not be thought as a well-defined constant.
Here, we use the exponent $\nu$ obtained in the temperature range 
under consideration only for a qualitative comparison with the experiments.
In Fig.~\ref{fig:paper4}, the exponent $\nu$
is shown as a function of $E_{\rm J}/E_{\rm C}$. We observed
the tendency that the exponent is approximately a linear function of 
$E_{\rm J}/E_{\rm C}$, and that the slope depend on $\alpha$. 

Let us compare the results obtained here with the experiment.
Experimentally, the temperature dependence of the zero-bias resistance
$R_0$ has been measured to characterize the superconductor-insulator 
transition; In Ref.~\cite{Yagi97} the system is assigned to an insulator
when $dR_0/dT<0$ at low temperatures, and otherwise to a superconductor.
Systematic analysis of zero-bias resistance, however, has not been performed 
for a single resistance-shunted Josephson junction~\cite{Yagi97,Pentila99}.
In the superconducting regime ($\alpha > 1$),
the zero-bias resistances in these experiments weakly depend
on the temperature, and do not show a clear power-law behavior. 
These experimental results disagree with the present theoretical calculation.
The origin of this disagreement is not known. 
We speculate that the environment effect through the leads may be 
important to determine the temperature dependence of $R_0$. 
Finally, we point out that the zero-bias resistance observed in 
one-dimensional Josephson arrays~\cite{Miyazaki02} obeys the
power-law behavior in the superconducting phase.
In this experiment, the exponent seems to depend linearly on 
$E_{\rm J}/E_{\rm C}$ and not on $\alpha$. Numerical calculation
of one-dimensional resistance-shunted Josephson arrays and comparison
with the experiment remain for a future problem.

In summary, we have studied the temperature dependence of the 
zero-bias resistance of a single resistance-shunted Josephson junction 
by means of quantum path-integral Monte Carlo method.
We reproduced the result predicted by Fisher and Zwerger 
even though $E_J/E_C \sim 1$. We have also shown
superconductor-insulator transition around $\alpha=R_Q/R_S=1$.
The temperature dependence of the zero-bias resistance showed a power-law-like
behavior in the temperature range under consideration.
In conflict with the renormalization analysis, 
the exponent depends on $E_{\rm J}/E_{\rm C}$.
These results have been compared with
the experiment on one-dimensional Josephson arrays.

\end{document}